\documentclass[prd,aps,eqsecnum,floatfix,nofootinbib,preprint,tightenlines]{revtex4}

\usepackage{latexsym}
\usepackage{graphicx}
\usepackage{multirow}
\usepackage[dvipsnames]{xcolor}

\def\bibi{\bibitem}
\def\floatcaption#1#2{ \caption{#2 \label{#1}} }

\def\a{\alpha}
\def\b{\beta}

\def\d{\delta}
\def\g{\gamma}

\def\m{\mu}
\def\n{\nu}

\def\p{\pi}                     
\def\r{\rho}                    
\def\t{\tau}

\def\D{\Delta}

\def\P{\Pi}

\def\cbo{{\,\raise-.15ex\Sc [\,}}                       

\def\svev#1{\left\langle #1\right\rangle}       

\def\ddt#1{{\buildrel {\hbox{\LARGE .\kern-2pt.}} \over {#1}}}

\def\Tr{{\rm Tr}\,}

\def\half{{1\over 2}}

\long \def \blockcomment #1\endcomment{}

\def\seef{{\it cf.\  }}

\def\hB{{\hat{B}}}

\begin{document}

\begin{center}
\vspace{10mm}
\begin{boldmath}
{\large\bf Determination of the NNLO low-energy constant $C_{93}$}\\[8mm]
\vspace{3ex}
\end{boldmath}
Maarten~Golterman,$^a$
 Kim~Maltman,$^{b,c}$ Santiago~Peris$^d$%
\\[0.1cm]
{\it
\null$^a$Department of Physics and Astronomy\\
San Francisco State University, San Francisco, CA 94132, USA\\
\null$^b$Department of Mathematics and Statistics\\
York University,  Toronto, ON Canada M3J~1P3\\
\null$^c$CSSM, University of Adelaide, Adelaide, SA~5005 Australia\\
\null$^d$Department of Physics and IFAE-BIST, Universitat Aut\`onoma 
de Barcelona\\
E-08193 Bellaterra, Barcelona, Spain}
\\[6mm]
{ABSTRACT}
\\[2mm]
\end{center}
\begin{quotation}
Experimental data from hadronic $\t$ decays allow for a precision
determination of the slope of the $I=1$ vacuum polarization
at zero momentum. We use this information to provide a value for the
next-to-next-to-leading order (NNLO) low-energy constant $C_{93}$ in
chiral perturbation theory.  The largest systematic error in
this determination results from the neglect of terms beyond
NNLO in the effective chiral Lagrangian, whose presence in the data
will, in general, make the effective $C_{93}$ determined in an NNLO
analysis mass dependent. We estimate the size of this effect by
using strange hadronic $\t$-decay data to perform an alternate
$C_{93}$ determination based on the slope of the strange vector
polarization at zero momentum, which differs from that of the
$I=1$ vector channel only through $SU(3)$ flavor-breaking effects.
We also comment on the impact of
such higher order effects on ChPT-based estimates for the hadronic
vacuum polarization contribution to the muon anomalous magnetic moment.
\end{quotation}

\vfill
\eject
\setcounter{footnote}{0}

\newpage
\section{\label{introduction} Introduction}
The spin $J=0+1$ polarization sums, $\Pi_{V/A;ud,us}^{0+1}$,
of the flavor $ud$ and $us$ vector ($V$) and axial vector ($A$) current
two-point functions of QCD have been calculated to two-loop order in
Chiral Perturbation Theory (ChPT) \cite{ABT}. It is therefore,
in principle, possible to provide estimates for low-energy constants
(LECs) appearing in these expressions at next-to-next-to-leading order
(NNLO) by comparing the relevant ChPT expressions to either dispersive
representations of the subtracted polarizations,
$\Pi_{V/A;ud,us}^{0+1,{\rm sub}}(Q^2)\equiv\Pi_{V;ud,us}^{0+1}(Q^2)
-\Pi_{V;ud,us}^{0+1}(0)$,
or inverse-moment finite-energy sum-rule (IMFESR) results for their
slopes at $Q^2\, =\, -q^2\, =\, -s=0$, both of which can be determined
from experimental data for the spectral functions in the $V$ and $A$ channels.

The LECs that appear in the ChPT expressions for
$\Pi_{V/A;ud,us}^{0+1}(Q^2)$ are the next-to-leading order (NLO)
LECs $L_9$ and $L_{10}$ and the NNLO LECs $C_{12}$, $C_{13}$,
$C_{61}$, $C_{62}$, $C_{80}$, $C_{81}$, $C_{87}$ and $C_{93}$ in the
$SU(3)$-flavor-symmetric limit, and, in addition, the NLO LEC $L_5$  in flavor-breaking contributions proportional
to $m_K^2-m_\p^2$.
In previous work \cite{L10,NNLO,NNLO2}, we provided
determinations of $L_{10}$ and the linear combinations $C_{12}-C_{61}+C_{80}$,
$C_{13}-C_{62}+C_{80}$, $C_{61}$ and $C_{87}$ using dispersive
and IMFESR results for the flavor $ud$ $V-A$ polarization and
flavor-breaking $ud-us$ $V$ and $V+A$ polarization combinations.
For the $ud$ $V-A$ polarization, both lattice results at unphysical
quark mass~\cite{Boyleetal} and physical-quark-mass results,
obtained using hadronic $\t$ decay data from OPAL \cite{OPAL} and
ALEPH \cite{ALEPH13} for the non-strange spectral functions, were
employed. The IMFESRs used to determine the $Q^2=0$ values of the
flavor-breaking $V$ and $V+A$ polarizations required, in addition,
strange hadronic $\t$-decay data from ALEPH~\cite{alephus99},
Belle~\cite{bellekspi,bellekspipi,belle3K} and
BaBar~\cite{babarkmpi0,babarkpipiallchg,babar3K}, together with
2014 HFAG strange branching fractions~\cite{HFAG2014}.

In the present paper, we consider the LEC $C_{93}$, which can be obtained
from a determination of the slope with respect to Euclidean
momentum-squared, $Q^2$, at $Q^2=0$, of the $V$ polarization using
the ALEPH data. $C_{93}$ is the only NNLO LEC appearing in the NNLO
representation of the subtracted polarizations
$\Pi_{ud,us}^{\rm sub}(Q^2)$.{\footnote{Since we will consider
only the spin $J=0+1$ $V$ case in this paper, we will drop the 
superscript $0+1$ and the subscript $V$ from now on.}} The $ud$ 
representation also depends on
the NLO LEC $L_9$ and the $us$ representation on the NLO LECs $L_5$ and $L_9$.
With $\Pi_{ud,us}^{\rm sub}(Q^2)$ both admitting once-subtracted dispersive
representations, $C_{93}$ can, in principle, be determined from
the experimental spectral data of either channel.
As in our previous work, we will take $L_5$ and $L_9$ from outside
sources \cite{MILC,BTL9}. In what follows, in addition to
$\Pi_{ud,us}^{\rm sub}$, we also consider the subtracted version of the
$V$ current polarization, $\Pi_\eta (Q^2)$ in the notation of Ref.~\cite{ABT},
associated with the neutral octet $V$ current
$(\bar{u}\gamma_\mu u +\bar{d}\gamma_\mu d-2\bar{s}\gamma_\mu s)/\sqrt{6}$.
$L_9$ and $C_{93}$ are the only NLO and NNLO LECs appearing in the NNLO
representation of $\Pi_\eta^{\rm sub}(Q^2)$.

Spectral functions (generically denoted $\rho (s)$) obtained
from hadronic $\t$ decays are, of course, limited to $s\le m_\t^2$.
This limits the radius, $s_0$, of the circular contour in the complex-$s$
plane used in $\tau$-based IMFESRs to $s_0\le m_\tau^2$. Dispersive
representations of the subtracted polarizations require the corresponding
$\rho (s)$ for all $s$. In the $ud$ channel, we will use a representation of
$\rho (s)$ above the $\t$ mass obtained from sum-rule-based fits
employing perturbation theory, augmented by a model for duality-violating
(resonance) effects, performed in Ref.~\cite{BGMOP14}. While this introduces
an assumption about the validity of this model into our extraction of
$C_{93}$, the low-$Q^2$ region from which $C_{93}$ is
determined is very insensitive to the details of this assumption.
Hence, we believe that the associated potential uncertainty
is far smaller than the systematic error due to the neglect of
orders beyond NNLO in ChPT.

Since we will employ ChPT to NNLO, the value of $C_{93}$ obtained
from the $ud$ $V$ channel analysis, which we denote by $C_{93}^{ud}$,
will have a residual mass dependence, originating from the effect of
beyond-NNLO loop and LEC contributions present in the data, but absent
from the NNLO representation of $\Pi_{ud}^{\rm sub}(Q^2)$. NNNLO loop
corrections have not been calculated, so an expanded NNNLO analysis is
not possible. We will, therefore, use flavor-breaking IMFESRs to obtain
an estimate for the slope of the difference of the $us$ and $ud$ $V$
polarizations, hence also of the slope of the $us$ $V$ polarization,
at $Q^2=0$. This latter result provides an alternate NNLO
$us$-$V$-channel-based determination, $C_{93}^{us}$, of $C_{93}$. The
difference between $C_{93}^{ud}$ and $C_{93}^{us}$ then provides an
estimate of the size of residual mass-dependent effects originating
from orders beyond NNLO.

This paper is organized as follows. In Sec.~\ref{theory} we summarize the
necessary theory, and in Sec.~\ref{C93} we present our central
$ud$-$V$-channel-based result for $C_{93}$, with the experimental
error coming from the ALEPH data. In Sec.~\ref{residual} we analyze the
$us-ud$ difference mentioned above, and obtain an estimate of the
systematic error due to the neglect of higher orders in ChPT.
In Sec.~\ref{muon} we comment on the use of NNLO ChPT for
estimates of the hadronic vacuum-polarization contribution to the
anomalous magnetic moment of the muon. We conclude with a
discussion of our results.

\section{\label{theory} Theory summary}
In this section we briefly summarize the necessary theory.

\subsection{\label{chpt} ChPT}
In the isospin limit, the expression for the subtracted vacuum polarization
$\P^{\rm sub}_{ud}(Q^2)$ was calculated to NNLO in ChPT in Ref.~\cite{ABT}.
As a function of Euclidean momentum-squared $Q^2$, it is given by
\begin{eqnarray}
\label{Pi1}
\P^{\rm sub}_{ud}(Q^2)&=&-8\hB(Q^2,m_\p^2)-4\hB(Q^2,m_K^2)\\
&&+\frac{16}{f_\p^2}\,L_9^r \,Q^2\left(2B(Q^2,m_\p^2)+B(Q^2,m_K^2)
\right)\nonumber\\
&&-\frac{4}{f_\p^2}\,Q^2\left(2B(Q^2,m_\p^2)+B(Q^2,m_K^2)
\right)^2+8C_{93}^rQ^2
\ ,
\nonumber
\end{eqnarray}
where $\hB(Q^2,m^2)=B(Q^2,m^2)-B(0,m^2)$ is the subtracted
standard, equal-mass, two-propagator, one-loop integral, with
\begin{eqnarray}
\label{B}
B(0,m^2)&=&\frac{1}{192\p^2}\left(1+\log{\frac{m^2}{\m^2}}\right)\ ,\\
\hB(Q^2,m^2)&=&\frac{1}{96\p^2}
\left(\left(\frac{4 m^2}{Q^2}+1\right)^{3/2}  \mbox{coth}^{-1}
   \sqrt{1+\frac{4m^2}{Q^2}}-
   \frac{4m^2}{Q^2}-{\frac{4}{3}}\right)\ ,\nonumber
\end{eqnarray}
and the low-energy constants (LECs) $L_9^r$ and $C_{93}^r$ are renormalized
at the scale $\m$, in the ``$\overline{MS}+1$'' scheme
employed in Ref.~\cite{ABT}.

From Eq.~(\ref{Pi1}) it is clear that $C_{93}^r$ can be determined from the
slope of $\P^{\rm sub}_{ud}(Q^2)$ at $Q^2=0$. Since we will only use
the explicit expression for $\P^{\rm sub}_{us}(Q^2)$ for a systematic
error estimate, we do not provide it here, but refer to Ref.~\cite{ABT} for the
full expression.\footnote{In the notation of Ref.~\cite{ABT},
$\Pi^{\rm sub}_{us}(Q^2) =\P^{(1)}_{VK}(-Q^2)+\P^{(0)}_{VK}(-Q^2)-
(\P^{(1)}_{VK}(0)+\P^{(0)}_{VK}(0))$.}

\subsection{\label{FB} Flavor-breaking sum rule}
The difference $\D\P(Q^2)$ of the $ud$ and $us$ spin $J=0+1$ $V$
unsubtracted two-point functions $\P_{ud}(Q^2)$ and $\P_{us}(Q^2)$
satisfies the flavor-breaking IMFESR
\begin{eqnarray}
\label{IMFESR}
\frac{d\D\P (Q^2)}{dq^2}\Big|_{q^2=0}\, =\,
-\frac{d\D\P(Q^2)}{dQ^2}\Big|_{Q^2=0}&=&
\int_{4m_\p^2}^{s_0} ds\,w_\t(s/s_0)\,\frac{\D\r(s)}{s^2}\\
&&+\frac{1}{2\p i}\oint_{|s|=s_0} ds\,w_\t(s/s_0)\,
\frac{\D\P(Q^2=-s)}{s^2}
\ ,\nonumber
\end{eqnarray}
where $q^2=-Q^2$, $w_\t(x)=(1-x)^2(1+2x)$,
and $\D\r(s)=\r_{ud}(s)-\r_{us}(s)$.
As long as we choose $s_0\le m_\t^2$,
the first integral on the right-hand side can be computed using
experimentally available spectral functions. We have used that
$w_\t(0)=1$ and $dw_\t(s/s_0)/ds|_{s=0}=0$.

As in other applications of FESRs, we will approximate $\D\P(s)$ in the
second integral by the operator product expansion
(OPE),\footnote{Mass-independent, purely perturbative contributions
cancel for the flavor-breaking combination considered here. The
leading, dimension-2, OPE contribution is thus of order $(m_s-m)^2/s$,
where $m$ is the $u$, $d$-averaged light quark mass.} and assume that
the contribution from duality violations to this sum rule are negligibly
small. In this case, this is reasonable because of the presence of a weight
function with a double pinch at $s=s_0$, as well as a further $1/s^2$
suppression of the contribution from higher-$s$ values to the integral.
This assumption can be tested for self-consistency by studying the
$s_0$ dependence of the right-hand side of Eq.~(\ref{IMFESR}). Because the
left-hand side is independent of $s_0$, the individually
$s_0$-dependent $ud$- and $us$-spectral integral and
OPE integral contributions should combine to produce a right-hand side
independent of $s_0$, within errors.

The sum rule~(\ref{IMFESR}) gives access to the difference of the
slopes of the subtracted polarizations $\Pi_{ud}^{\rm sub}(Q^2)$ and
$\Pi_{us}^{\rm sub}(Q^2)$ at $Q^2=0$. This, together with the independent
dispersive determination of the slope of $\Pi_{ud}^{\rm sub}(Q^2)$, yields
the value of the slope of $\Pi_{us}^{\rm sub}(Q^2)$ at $Q^2=0$. The NNLO
expression for this slope provides the alternate determination,
$C^{us}_{93}$, of $C_{93}$ already introduced above.

Of course, since LECs are, by definition, mass independent,
the NNLO analysis results $C_{93}^{ud}$ and $C_{93}^{us}$ should
be the same, provided NNNLO and higher order contributions are
negligible. The experimental data used in their determination, however,
know about the existence of higher orders in ChPT, and if these are
not, in fact, negligible, we expect the two values to be different.
The numerical difference provides an indication of the size of
higher-order, residual mass-dependent effects.

In ChPT, the leading mass-dependent corrections to the slopes of the
$V$ polarizations at $Q^2=0$ result from NNNLO operators having
a single insertion of the chiral mass operator ($\chi_+$ in the notation of
Ref.~\cite{bce99}). A two-trace NNNLO operator of this form in which $\chi_+$
appears through the factor $\Tr (\chi_+)$ produces an
$SU(3)$-flavor-symmetric contribution proportional to
$2m_K^2+m_\pi^2$ to the slopes of all of
the $ud$, $us$ and $\eta$ $V$ channel polarizations at $Q^2=0$. A
single-trace NNNLO operator containing one factor of $\chi_+$, similarly,
produces contributions proportional to $m_\pi^2$, $m_K^2$ and
$m_\eta^2={\frac{4}{3}}m_K^2-{\frac{1}{3}}m_\pi^2$, respectively, to those
same slopes.\footnote{Here we use the tree-level relations between
quark and meson masses.} To take these effects into account, we introduce
two NNNLO LECs, $\d C_{93}^{(1)}$ and $\d C_{93}^{(2)}$,
where the $(1),\, (2)$ superscripts indicates the number of trace factors
in the accompanying operators, normalized such that they produce
mass-dependent contributions
\begin{eqnarray}
\label{C93udus}
C_{93}^{ud}&=&C_{93}^r+\d C_{93}^{(2)}(2m_K^2+m_\p^2)
+\d C_{93}^{(1)}m_\p^2\ ,\\
C_{93}^{us}&=&C_{93}^r+\d C_{93}^{(2)}(2m_K^2+m_\p^2)
+\d C_{93}^{(1)}m_K^2\ ,\nonumber\\
C_{93}^\eta&=&C_{93}^r+\d C_{93}^{(2)}(2m_K^2+m_\p^2)
+\d C_{93}^{(1)}\left(\frac{4}{3}\,m_K^2-\frac{1}{3}\,m_\p^2\right)
\ .\nonumber
\end{eqnarray}

Only the first of the new NNNLO LECs, $\delta C_{93}^{(1)}$,
contributes to the difference $C^{ud}_{93}-C^{us}_{93}$ at NNNLO,
and thus if, guided by what is found at NNLO, we assume
NNNLO LEC contributions will dominate loop contributions also
at NNNLO, the sum rule~(\ref{IMFESR}) will give us an estimate of
$\d C_{93}^{(1)}$.
The observation that $\d C_{93}^{(2)}$ is suppressed in large $N_c$
then leads to two expectations: one that $C^{ud}_{93}$ should be
much closer to the true, mass-independent $C_{93}$ than $C_{93}^{us}$;
the other that the difference $C^{ud}_{93}-C^{us}_{93}$ should give a
reasonably conservative estimate of the systematic error associated
with the neglect of contributions beyond NNLO in ChPT. We emphasize
that loop contributions at NNNLO will also produce mass-dependent
contributions to the slopes of $\P^{\rm sub}_{ud}(Q^2)$ and
$\P^{\rm sub}_{us}(Q^2)$, and thus that this estimate relies on the
assumption, also made in the rest of this paper, that
such mass-dependent higher-order loop contributions are small
compared to the LEC contributions, at the scale $\m=0.77$~GeV 
in the $\overline{MS}+1$ scheme we will use in this paper.

We note in closing this section that the higher-order,
mass-dependent effects discussed above are also included
in the phenomenological approach of Ref.~\cite{BR16}, where NNLO and
higher LEC contributions are modelled by replacing the NNLO
LEC contributions proportional to $C_{93}^r$ in the expressions
for the subtracted polarizations with the corresponding full
vector-meson dominance (VMD) contributions obtained using
$\rho$ and $\phi$ masses in the VMD expressions for the $I=1$
and strange current channels, respectively. The chiral-limit
part of the vector meson mass in this approach
produces quark-mass-independent contributions which, in the chiral
expansion, would be parametrized by a tower of NNLO and higher
LECs, including $C_{93}^r$ and the NNNLO LEC $C^r$ introduced in
Ref.~\cite{NNLO} (which produces a common $SU(3)$-flavor-symmetric contribution
$C^rQ^4$ to the subtracted $V$ polarizations we consider in this paper).
The quark-mass-dependent parts of the different vector meson masses
used in the different $V$ channels, similarly, generate contributions
which would be parametrized by $\delta C_{93}^{(1)}$,
$\delta C_{93}^{(2)}$, and yet higher-order LECs. The VMD extension
of the NNLO results contains only contributions analytic in $Q^2$
in the low-$Q^2$ region and hence also neglects NNNLO and higher loop
contributions.

\begin{boldmath}
\section{\label{C93} $C_{93}$ from ALEPH data}
\end{boldmath}

The once-subtracted $ud$  $V$ polarization $\P^{\rm sub}_{ud}(Q^2)$
can be defined in terms of the corresponding spectral function
$\r_{ud}(s)$ as a function of the Euclidean momentum-squared
$Q^2$ by the dispersion relation
\begin{equation}
\label{disprel}
\P^{\rm sub}_{ud}(Q^2)=-Q^2\int_{4m_\p^2}^\infty ds\,
\frac{\r_{ud}(s)}{s(s+Q^2)}\ .
\end{equation}
For $s<m_\t^2$, we can use the experimental spectral function provided by
Ref.~\cite{ALEPH13}, but for $s>m_\tau^2$ we will need a theoretical
representation, with parameters fit from the data in the region below
$m_\t^2$. We follow the procedure employed in
Refs.~\cite{L10,NNLO2,gminus2chiral}, using the fitted version
of the theoretical representation obtained starting from
the rescaled version of the data for the ALEPH $ud$ spectral function,
and following the procedure described in detail in Ref.~\cite{BGMOP14}.
The theoretical representation is the sum of the QCD perturbation
theory (PT) expression $\r_{ud,\rm PT}(s)$ and a ``duality-violating'' (DV)
part $\r_{ud,\rm DV}(s)$ representing the effects of resonances, with
the {\it ansatz}
\begin{equation}
\label{DV}
\r_{ud,\rm DV}(s)=e^{-\d_V-\g_V s}\sin{(\a_V+\b_V s)}
\end{equation}
used for the DV part.
The perturbative expression is known to order $\a_s^4$ \cite{PT}.
Fits to the weighted integrals of the ALEPH data determining
the parameters $\a_s$, $\a_V$, $\b_V$, $\g_V$ and $\d_V$ have been
performed in Ref.~\cite{BGMOP14}, with a focus on the
high-precision determination of $\a_s$ from hadronic $\t$ decays.
We will use the values obtained from the FOPT $s_{\rm switch}
=s_{min}=1.55$~GeV$^2$ fit of Table~1 of Ref.~\cite{BGMOP14},
\begin{eqnarray}
\label{param}
\a_s(m_\t^2)&=&0.295(10)     \ ,\\
\a_V&=&-2.43(94)     \ ,\nonumber\\
\b_V&=&4.32(48)\ \mbox{GeV}^{-2}     \ ,\nonumber\\
\g_V&=&0.62(29)\ \mbox{GeV}^{-2}     \ ,\nonumber\\
\d_V&=&3.50(50)     \ .\nonumber
\end{eqnarray}
The matches between the data and theory representations of both
the weighted spectral integrals and the spectral function in the
window used in performing the fits are excellent, and there
is no discernible effect on $\P^{\rm sub}_{ud}(Q^2)$ for the values
of $Q^2$ smaller than $0.2$~GeV$^2$ of interest in the comparison to ChPT
if we vary the point at which we switch from the experimental
to the theoretical version of $\rho_{ud}(s)$ within this fit window,
use the results of a CIPT instead of an  FOPT fit, or employ parameter
values from one of the other optimal fits in Ref.~\cite{BGMOP14}.
Results for $\P^{\rm sub}_{ud}(Q^2)$ in the region below $Q^2=0.2$~GeV$^2$,
at intervals of $0.01$~GeV$^2$, are shown in Fig.~\ref{PiVdata}. The
errors shown are fully correlated, taking into account, in particular,
correlations between the parameters of Eq.~(\ref{param}) and the data. We
emphasize again that systematic effects due to the use of the
{\it ansatz}~(\ref{DV}) can be assumed to be small compared
to systematic effects due to the neglect of higher orders in ChPT.
\begin{figure}[t]
\vspace*{4ex}
\begin{center}
\includegraphics*[width=14cm]{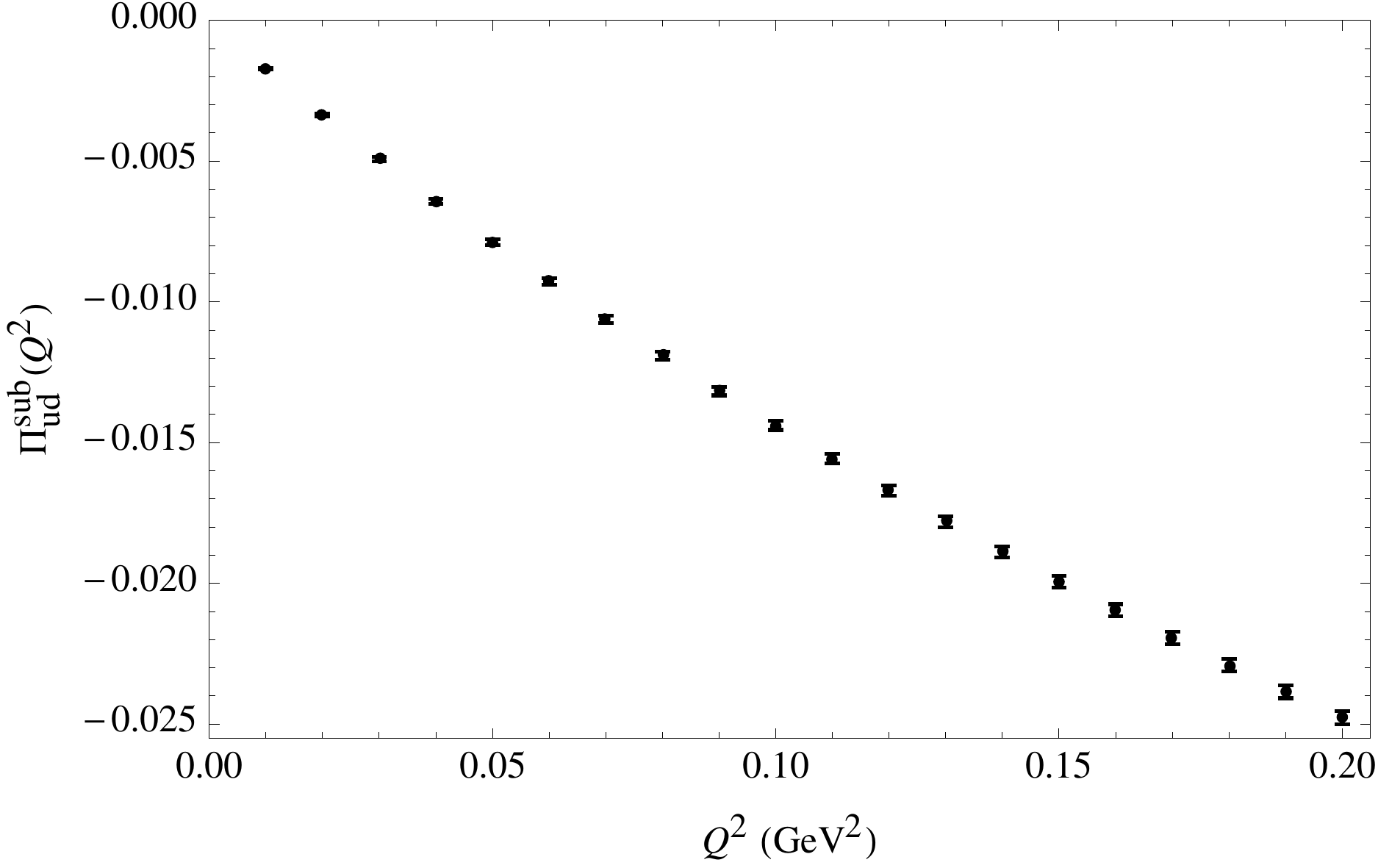}
\end{center}
\begin{quotation}
\floatcaption{PiVdata}%
{\it $\P^{\rm sub}_{ud}(Q^2)$ as a function of $Q^2$
constructed from ALEPH data as explained in Sec.~\ref{C93}.}
\end{quotation}
\vspace*{-4ex}
\end{figure}

It follows from Eq.~(\ref{Pi1}) that the slope of $\P^{\rm sub}_{ud}(Q^2)$
at $Q^2=0$ is a linear combination of the LECs $L^r_9$ and $C^r_{93}$.
We will use
\begin{eqnarray}
\label{masses}
m_\p&=&139.57~\mbox{MeV}\ ,\\
m_K&=&495.65~\mbox{MeV}\ ,\nonumber\\
f_\p&=&92.21~\mbox{MeV}\ .\nonumber
\end{eqnarray}
The errors on these values are so small that they can be
ignored in the computation of the error on $C^r_{93}$. We also use
the value \cite{BTL9}
\begin{equation}
\label{L9}
L^r_9(\m=0.77\ \mbox{GeV})=0.00593(43)\ .
\end{equation}
With these inputs, the NNLO representation of the slope, Eq.~(\ref{Pi1}),
becomes
\begin{equation}
\label{slopeudNNLO}
\frac{d\P^{\rm sub}_{ud}(Q^2)}{dQ^2}\Big|_{Q^2=0}=
\left( -0.02253-0.00291-0.02775(201)\right)\ {\rm GeV}^{-2} +8C_{93}^{ud}\, ,
\end{equation}
where the first term is the NLO contribution, the second the NNLO loop
contribution involving only LO vertices and the third the NNLO loop
contribution proportional to $L_9^r$. The slope obtained from the results for
$\P^{\rm sub}_{ud}(Q^2)$ shown in Fig.~\ref{PiVdata},\footnote{The
error on this value is based on propagation of the full covariance matrix.}
\begin{equation}
\label{slope}
\frac{d\P^{\rm sub}_{ud}(Q^2)}{dQ^2}\Big|_{Q^2=0}
=-0.17608\pm 0.00291\ \mbox{GeV}^{-2}\, ,
\end{equation}
then yields, for the (potentially mass-dependent) $ud$ channel effective LEC
$C_{93}^{ud}$, the result
\begin{equation}
\label{C93ud}
C_{93}^{ud}(\m=0.77\ \mbox{GeV})=-0.01536 \pm 0.00036\pm
0.00025 \ \mbox{GeV}^{-2}\ ,
\end{equation}
where the first error comes from the error in the slope, and the second
from the error in $L_9^r$. As one can see, the result in (\ref{slopeudNNLO}) 
is dominated by the contribution from $C_{93}^{ud}$. Residual mass-dependent
effects causing $C_{93}^{ud}$ to, in principle, differ from $C_{93}^r$
remain to be estimated.

\vskip0.8cm
\section{\label{residual} Estimate of residual mass dependence}

The value for $C_{93}$ obtained in Eq.~(\ref{C93ud}) appears to have a very
small error, but this is misleading. There are, in fact, other small
errors which we neglected in this result, for instance due to the use
of our {\it ansatz}~(\ref{DV}) and isospin breaking. However, as mentioned
already above, there is also a systematic error due to the neglect of
orders in ChPT beyond NNLO, which is likely to be more important,
and which we address in this section. We do so by using the IMFESR~(\ref{IMFESR}) to determine the difference in the slopes at $Q^2=0$ of
$\P^{\rm sub}_{ud}(Q^2)$ and $\P^{\rm sub}_{us}(Q^2)$, and, from this,
using the result for $d\P^{\rm sub}_{ud}(Q^2)/dQ^2|_{Q^2=0}$ from
Eq.~(\ref{slope}), determine $d\P^{\rm sub}_{us}(Q^2)/dQ^2|_{Q^2=0}$, whose
NNLO representation then provides us with an alternate, $us$-channel
determination of $C_{93}$, denoted $C_{93}^{us}$. $C_{93}^{ud}$ and
$C_{93}^{us}$ should be equal within errors if contributions beyond
NNLO are negligible. As it turns out, they are not, and this allows
us to estimate the effect of the neglect of higher
order contributions. While it is difficult to convert
this estimate into a reliable systematic error, it will be clear that
this is the dominant uncertainty in our result for $C_{93}^r$.

In order to evaluate the right-hand side of Eq.~(\ref{IMFESR}), we
will need the $ud$ and $us$ spectral functions, as well as the
dimension-2 and dimension-4 terms in the OPE.

The $ud$ spectral function is the same as that used to construct
$\P^{\rm sub}_{ud}(Q^2)$ above; we refer again to Ref.~\cite{BGMOP14}
for a more detailed discussion (\seef\ Sec.~III.A, in particular).

The $us$ spectral function we will use is constructed as a sum over
exclusive mode contributions, as in Ref.~\cite{NNLO}, to which we refer
for a detailed discussion (\seef\ Sec.~III.C, in particular).
All 2014 HFAG inputs used previously have been updated to reflect
current 2016 HFAG values \cite{HFAG16}.

One additional issue to consider is the choice of $K\p$
branching fractions. These provide the overall normalization used to
convert the unit-normalized Belle experimental distribution~\cite{bellekspi}
to the actual, physically normalized $K\pi$ contribution to the $us$
spectral function. The first normalization is the one provided by
HFAG \cite{HFAG16},
\begin{eqnarray}
\label{HFAGnorm}
B[K^-\pi^0 ]&=&0.00433(15)\ ,\\
B[\bar{K}^0\pi^-]&=&0.00839(14)\ ,\nonumber
\end{eqnarray}
the errors of which are essentially uncorrelated, yielding
a 2-mode $K\pi$ branching fraction sum $0.01271(21)$. The dispersive
study of Ref.~\cite{ACLP2013}, however, finds clear tension between such
branching fraction values, $K_{\ell 3}$ results and dispersive
constraints on the $K\pi$ form factors. The analysis of Ref.~\cite{ACLP2013}
yields slightly higher expectations for these branching fractions,
\begin{eqnarray}
\label{ACLPnorm}
B[K^-\pi^0 ]&=&0.00471(18)\ ,\\
B[\bar{K}^0\pi^-]&=&0.00857(30)\ ,\nonumber
\end{eqnarray}
this time with the errors essentially 100\% correlated and hence
a 2-mode $K\pi$ branching fraction sum $0.01327(48)$. We consider
both possibilities in our analysis; the associated $ud-us$ slope
uncertainty is found to be about half the size of the error
induced by other experimental uncertainties.

We treat the OPE in the same way as in Ref.~\cite{NNLO}, and
refer to Sec.~III.B of Ref.~\cite{NNLO} for the explicit expressions.
We will use the input parameters\footnote{Because the OPE contribution
to Eq.~(\ref{IMFESR}) is so small, it does not matter whether one uses the value
for $\a_s(m_\t^2)$ given below, or the one given in Eq.~(\ref{param}).}
\begin{eqnarray}
\label{OPEpar}
\a_s(m_\t^2)&=&0.3155(90)\ \qquad (\mbox{converted\ from\ Ref.~\cite{PDG}})\ ,\\
m_s(2\ \mbox{GeV})&=&93.9(1.1)\ \mbox{MeV}\ \qquad(\mbox{Ref.~\cite{FLAG16}})\ ,
\nonumber\\
m_\t&=&1.77686(12)\ \mbox{MeV}\qquad(\mbox{Ref.~\cite{PDG}})\ ,\nonumber\\
\svev{{\overline{s}}s}/\svev{\overline{u}u}&=&1.08(16)
\qquad(\mbox{Ref.~\cite{HPQCD}})\ ,
\nonumber\\
B_e&=&0.17815(23)\qquad(\mbox{Ref.~\cite{HFAG16}})\ ,
\nonumber\\
V_{ud}&=&0.97417(21)\qquad(\mbox{Ref.~\cite{HT14}})\ ,
\nonumber\\
V_{us}&=&0.22582(91)\qquad(\mbox{3-family\ unitarity})\ ,
\nonumber\\
S_{EW}&=&1.0201(3)\qquad(\mbox{Ref.~\cite{Erler}})\ ,\nonumber
\end{eqnarray}
where $\svev{\overline{u}u}$ is in the isospin limit, and its value is
obtained from the GMOR relation. We find that the OPE contribution to the
right-hand side of Eq.~(\ref{IMFESR}) is less than 1.6\% of the total for
$s_0=2.15$~GeV$^2$, and decreases for larger values of $s_0$.

Very good $s_0$-stability is observed for the slope obtained
from this analysis. This is illustrated, for the ACLP choice of the $K\pi$
normalization, in Figure~\ref{imfesrbreakdown}. The figure shows the
individual terms (OPE integral, $ud$ spectral integral and $us$
spectral integral) appearing on the right-hand side of Eq.~(\ref{IMFESR}), together
with the $ud$-$us$+OPE combination which determines
$-{\frac{d\Delta\Pi(Q^2)}{dQ^2}}\vert_{Q^2=0}$, all as a function of $s_0$.
The corresponding results for the HFAG $K\pi$ normalization choice
are essentially identical, and hence not shown explicitly. The excellent
$s_0$-stability provides a self-consistency check on our neglect of
duality violations employing the IMFESR~(\ref{IMFESR}), and confirms 
the very minor role played by the OPE.
\begin{figure}[t]
\vspace*{4ex}
\begin{center}
\rotatebox{0}{\mbox{
\begin{minipage}[t]{10.1cm}
\includegraphics*[width=12cm]{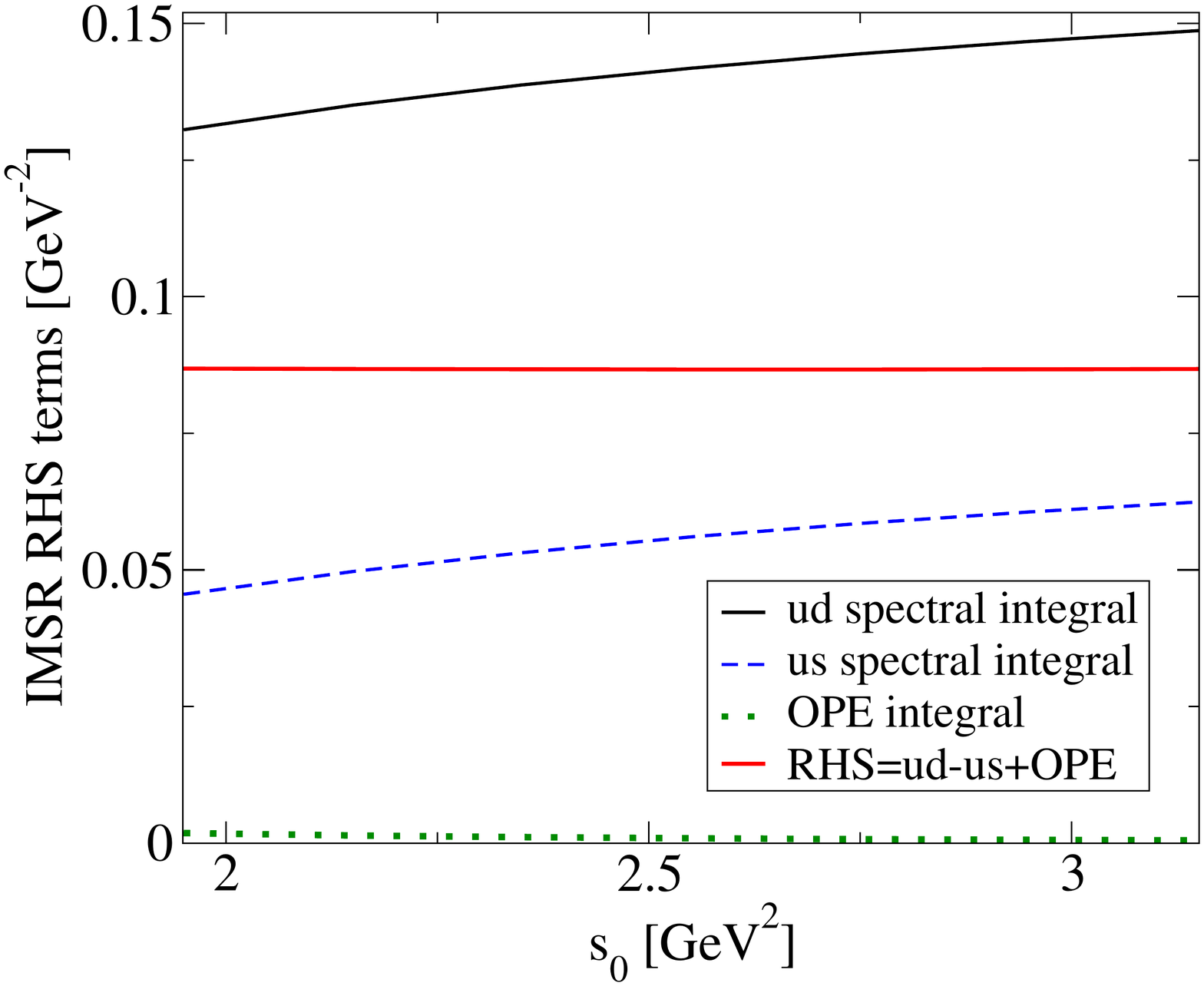}
\end{minipage}}}
\end{center}
\begin{quotation}
\floatcaption{imfesrbreakdown}%
{\it Contributions to the right-hand side (RHS) of the 
IMFESR~(\ref{IMFESR}) and the
resulting $ud$-$us$+OPE sum, as a function of $s_0$. The $us$ spectral
integrals are those obtained using the ACLP branching-fraction normalization of
the $K\pi$ distribution.}
\end{quotation}
\vspace*{-4ex}
\end{figure}

Using the $K\p$ branching-fraction normalization of Eq.~(\ref{HFAGnorm}),
and quoting the $s_0=m_\tau^2$ result to be specific,
we find a slope for the $ud-us$ difference
\begin{equation}
\label{slopedelta1}
\frac{d\D\P(Q^2)}{dQ^2}\Big|_{Q^2=0}=-0.0894(35)\ \mbox{GeV}^{-2}\, ,
\end{equation}
which yields
\begin{equation}
\label{slopeus1}
\frac{d\P^{\rm sub}_{us}(Q^2)}{dQ^2}\Big|_{Q^2=0}=
-0.0867(46)\ \mbox{GeV}^{-2}\,
\end{equation}
for the slope in the $us$ channel.
The corresponding NNLO representation, with Eq.~(\ref{masses}) as input, is
\begin{equation}
\label{slopeusnnlogenL59}
\frac{d\P^{\rm sub}_{us}(Q^2)}{dQ^2}\Big|_{Q^2=0}=
\left( -0.000868-0.004740-0.6836L^r_5-0.5419L^r_9\right)\ {\rm GeV}^{-2}
+8C^{us}_{93}\, ,
\end{equation}
with the first term the NLO contribution, the second the NNLO loop
contribution involving only LO vertices and the third and fourth the
NNLO one-loop contributions with a single NLO vertex. Using \cite{FLAG16}
\begin{equation}
\label{L5}
L_5^r(\mu=0.77\ \mbox{GeV})=0.00119(25)\ ,
\end{equation}
and Eq.~(\ref{L9}) for $L_9^r$, we have, adding the first two terms on the
right-hand side of Eq.~(\ref{slopeusnnlogenL59}),
\begin{equation}
\label{slopeusNNLO}
\frac{d\P^{\rm sub}_{ud}(Q^2)}{dQ^2}\Big|_{Q^2=0}=
\left(-0.005606-0.000814(171)-0.003214(230)\right)\ {\rm GeV}^{-2}
+8C_{93}^{us}\, ,
\end{equation}
and hence
\begin{equation}
\label{C93us1}
C^{us}_{93}(\mu=0.77\ \mbox{GeV})=-0.00963(58)\ \mbox{GeV}^{-2}\ ,
\end{equation}
where the error is dominated by the experimental error on the
$ud-us$ slope. Again, as in the ud channel, the slope~(\ref{slopeusNNLO}) is dominated by the contribution from the 
LEC $C^{us}_{93}$.

Using, instead, the $K\p$ branching-fraction normalization
of Eq.~(\ref{ACLPnorm}), we find a slope for the $ud-us$ difference
\begin{equation}
\label{slopedelta2}
\frac{d\D\P(Q^2)}{dQ^2}\Big|_{Q^2=0}=-0.0868(40)\ \mbox{GeV}^{-2}
\end{equation}
at $s_0=m_\t^2$, yielding for the slope in the $us$ channel
\begin{equation}
\label{slopeus2}
\frac{d\P^{\rm sub}_{us}(Q^2)}{dQ^2}\Big|_{Q^2=0}
=-0.0893(49)\ \mbox{GeV}^{-2}\ ,
\end{equation}
and the result
\begin{equation}
\label{C93us2}
C^{us}_{93}(\mu=0.77\ \mbox{GeV})=-0.00996(61)\ \mbox{GeV}^{-2}\ .
\end{equation}
The values~(\ref{C93us1}) and~(\ref{C93us2}) are consistent within errors.
Comparing these results with that in Eq.~(\ref{C93ud}) shows
the existence of significant residual mass-dependent effects.
Taking the average of the values~(\ref{C93us1}) and~(\ref{C93us2}) yields
\begin{equation}
\label{DeltaC93}
\frac{C^{ud}_{93}-C^{us}_{93}}{C^{ud}_{93}}\Big|_{\mu=0.77\ \rm GeV}=0.36(4)\ .
\end{equation}
The size of this difference is consistent with the expectation for an $SU(3)$
breaking effect. Finally, from
\begin{equation}
\label{der}
\frac{d\D\P(Q^2)}{dQ^2}\Big|_{Q^2=0}=\left(-0.019832+0.6836L_5^r
-4.1376L_9^r\right)\ {\rm GeV}^{-2}-8\d C_{93}^{(1)}(m_K^2-m_\p^2)\ ,
\end{equation}
we find
\begin{equation}
\label{C931}
\d C_{93}^{(1)}(m_K^2-m_\p^2)=
\left\{\begin{array}{ll}
0.00573(49)\ \mbox{GeV}^{-2} &\mbox{(HFAG)}\ ,\\
0.00540(55)\ \mbox{GeV}^{-2} &\mbox{(ACLP)}\ ,
\end{array}\right.
\end{equation}
for the $K\p$ branching-fraction normalizations of Ref.~\cite{HFAG16} and
Ref.~\cite{ACLP2013}, respectively.

\section{\label{muon} ChPT and the muon anomalous magnetic moment}
The lowest-order hadronic contribution to the muon anomalous magnetic
moment is given by an integral over $Q^2$ of the hadronic vacuum
polarization times a weight that causes about 90\% of the integral to
correspond to the integral between $Q^2=0$ and $Q^2=0.2$~GeV$^2$. One
may thus hope that ChPT can be used to constrain the low-momentum part
of this integral \cite{BR16,AB07,DJ10,GMP17}. In particular, since it is
difficult to compute the quark-disconnected part of the hadronic vacuum
polarization on the lattice \cite{Mainz14,hpqcddisc,RBCdisc,BMWdisc}, ChPT has
been used to estimate the size of the disconnected contribution relative
to the connected contribution \cite{BR16,DJ10}.

In Ref.~\cite{RBCdisc}, the disconnected part has been computed on the
lattice. In this analysis, an estimate of the systematic uncertainty
associated with the inability to accurately resolve the disconnected
signal at large Euclidean times was achieved by considering the Fourier
transform $\P_{uu-ss,dd-ss}(Q^2)$ of $\langle 0\vert T\left[ V^\m_{uu-ss}(x)
V^\n_{dd-ss}(0)\right]\vert 0\rangle$, which (in the isospin limit) is
equal to 9 times the sum of the connected strange and the full disconnected
contributions to the electromagnetic vacuum polarization. A physical
model for the large-time disconnected contribution was then obtained by
subtracting from a fitted two-exponential representation of the
strange-connected-plus-full-disconnected sum the well-determined strange
connected contribution. Though restricted in Ref.~\cite{RBCdisc} to an
investigation of the behavior of the disconnected contribution at
large Euclidean times, this strategy is, in principle, usable more
generally. Thus, were a reliable continuum representation of the
strange-connected-plus-full-disconnected sum to be available, the
disconnected contribution to the electromagnetic polarization
could be obtained from this simply by subtracting lattice results for
the strange connected contribution, which, for example, has been accurately
determined in~Refs.~\cite{hpqcdstrange,rbcukqcdstrange}. The hope is
that ChPT might provide such a reliable continuum representation,
at least in the low-$Q^2$ region. To see that this might indeed be
possible, note that one has, in terms of
the $I=1$ and $SU(3)$-octet vacuum polarizations $\P^{(1)}_{V\p}$
and $\P^{(1)}_{V\eta}$ of Ref.~\cite{ABT},
\begin{equation}
\label{uuddss}
\P_{uu-ss,dd-ss}(Q^2)=-\half\,\P^{(1)}_{V\p}(Q^2)
+\frac{3}{2}\,\P^{(1)}_{V\eta}(Q^2)\ .
\end{equation}
The results of Ref.~\cite{ABT} thus provide an NNLO representation
of $\P_{uu-ss,dd-ss}(Q^2)$. From Eq.~(\ref{C93udus}), the
``effective'' $C_{93}$ contribution, including NNNLO residual mass effects,
to $\frac{1}{9}\P_{uu-ss,dd-ss}(Q^2)$ is equal to
\begin{eqnarray}
\label{uuddssresmass}
\frac{8}{9}Q^2 C^{\rm eff}_{93}\equiv
&=&\frac{8}{9}Q^2\left(C^r_{93}+\d C_{93}^{(1)}(2m_K^2-m_\p^2)
+\d C_{93}^{(2)}(2m_K^2+m_\p^2)\right)\\
&=&\frac{8}{9}Q^2\left(C^{ud}_{93}+2\, \d C_{93}^{(1)}(m_K^2-m_\p^2)\right)
\ .\nonumber
\end{eqnarray}
Using Eqs.~(\ref{C93ud}) and~(\ref{C931}), we find a significant cancellation
between the $C^{ud}_{93}$ and the $\d C_{93}^{(1)}$ contributions in
(\ref{uuddssresmass}) resulting in
\begin{equation}
\label{C93eff}
C^{\rm eff}_{93}=\left\{
\begin{array}{ll}
-0.0039(11)\ \mbox{GeV}^{-2} &\mbox{(HFAG)}\ ,\\
-0.0046(12)\ \mbox{GeV}^{-2} &\mbox{(ACLP)}\ .
\end{array}\right.
\end{equation}
The two estimates are consistent within errors,
but very different from our best estimate for the true
value of $C_{93}^r$, given in Eq.~(\ref{C93ud}). The strong cancellation
between the $C^{ud}_{93}$ and the $\d C_{93}^{(1)}$ contributions
produces a result for the slope of ${\frac{1}{9}}\left[
-{\frac{1}{2}}\Pi_{V\pi}^{(1)} +{\frac{3}{2}}\Pi_{V\eta}^{(1)}\right]$
much less strongly dominated by the effective NNLO LEC combination
$C_{93}^{ud}+2(m_K^2-m_\pi^2)\delta C_{93}^{(1)}$ than is the case
for the slopes of either of the individual terms entering the difference.
Explicitly, one finds for the slope of this combination
\begin{eqnarray}
\label{slopenum}
&&\left[ 0.00082+0.00016+0.00189(14)\right]\ {\rm GeV}^{-2}
+{\frac{8}{9}}\,C_{93}^{\rm eff}\nonumber\\
&&=0.00287(14)\ {\rm GeV}^{-2}+{\frac{8}{9}}\,C_{93}^{\rm eff}\ ,
\end{eqnarray}
where the first three terms in the first line are the NLO
contribution, the NNLO loop contribution with only LO vertices, and the
NNLO loop contribution proportional to $L_9^r$, respectively, all at
$\mu =0.77$ GeV. The results given in Eq.~(\ref{C93eff}) yield for the last
contribution, $8C_{93}^{\rm eff}/9$, the values $-0.0035(9)$ and
$-0.0041(10)$ GeV$^{-2}$, for the HFAG and ACLP $K\pi$ normalization choices,
respectively.\footnote{It is worth noting that, though the VMD estimates
for $C_{93}^{ud}$ and $C_{93}^{us}$ differ by $\sim 10-30\%$ from the
corresponding dispersive and IMFESR determinations (see below
for details), the VMD estimate for $C_{93}^{\rm eff}$ works rather
well. Explicitly, with $f_{{\rm EM},V}$ the vector meson decay constants,
$\langle 0\vert J_\mu^{\rm EM}\vert V(q)\rangle
= g_{{\rm EM},V} m_V\epsilon_\mu (q) = f_{{\rm EM},V} m_V^2\epsilon_\mu (q)$,
one finds the VMD expectation
\begin{equation}
\label{vmd}
{\frac{8}{9}}\,C_{93}^{\rm eff}={\frac{1}{9}}
{\frac{f^2_{{\rm EM},\rho}}{m_\rho^2}}\, -\,
{\frac{f^2_{{\rm EM},\omega}}{m_\omega^2}}\, -\,
{\frac{f^2_{{\rm EM},\phi}}{m_\phi^2}}\ .
\nonumber
\end{equation}
With PDG values for the masses and $V\rightarrow e^+ e^-$ decay widths,
$g_{{\rm EM},\rho} = 156.4$ MeV, $g_{{\rm EM},\omega} = 46.6$ MeV and
$g_{{\rm EM},\phi} = 75.9$ MeV, the VMD estimate yields
$C_{93}^{{\rm eff}}\, =\, -0.0041\ {\rm GeV}^{-2}$, in
good agreement with the results of Eq.~(\ref{C93eff}).}
These are only slightly larger in magnitude than the sum of the NLO
and other NNLO contributions. In contrast, for
${\frac{d\Pi_{V\pi}^{(1)}(Q^2)}{dQ^2}}\Big|_{Q^2=0}$, the results
of Eq.~(\ref{slopeudNNLO}) show a $\mu =0.77$ GeV NNLO contribution proportional
to $C_{93}^{ud}$ a factor of $\sim 5.5$ larger than the corresponding
NLO contribution and $\sim 4.0$ larger than the remaining NNLO contributions.
The slope ${\frac{d\Pi_{V\eta}^{(1)}(Q^2)}{dQ^2}}\Big|_{Q^2=0}$ is even more
strongly dominated by the effective NNLO LEC contribution, with
\begin{equation}
{\frac{d\Pi_{V\eta}^{(1)}(Q^2)}{dQ^2}}\Big|_{Q^2=0}=
\left( -0.00258-0.00002+0.00210(15)\right)\ {\rm GeV}^{-2}
+8C_{93}^\eta \, ,
\end{equation}
where $C_{93}^\eta \equiv C_{93}^{ud}+{\frac{4}{3}}\delta C_{93}^{(1)}
(m_K^2-m_\p^2)$,
the first three terms have the same meaning as in Eq.~(\ref{slopeudNNLO})
and, with the results for $C_{93}^{ud}$ and $\delta C_{93}^{(1)}$ given
above, $8C_{93}^\eta =-0.0653(56)\ {\rm GeV}^{-2}$ and
$-0.0618(53)\ {\rm GeV}^{-2}$, for the ACLP and HFAG $K\pi$ normalization
cases, respectively. Moreover, in contrast to the $\Pi_{V\pi}^{(1)}$ and
$\Pi_{V\eta}^{(1)}$ cases, where the effective NNLO LEC contributions
have the same signs as the NLO and remaining NNLO contributions, the
effective NNLO LEC contribution to the slope in Eq.~(\ref{slopenum}) has
the opposite sign, leading to further cancellation between the effective
NNLO LEC and other contributions. The final values for the slope of the
${\frac{1}{9}}\left[ -{\frac{1}{2}}\Pi_{V\pi}^{(1)}
+{\frac{3}{2}}\Pi_{V\eta}^{(1)}\right]$ combination, $-0.0006(10)$
GeV$^{-2}$ and $-0.012(11)$ GeV$^{-2}$ for the HFAG and ACLP $K\pi$
normalization choices, respectively, thus show a further factor of
$3$ to $6$ reduction relative to the already reduced effective NNLO LEC
contributions. This raises the question of how safe it is to neglect NNNLO
and higher loop contributions for this particular combination.\footnote{
Significant cancellation in the LEC contributions is, in fact,
expected. If one neglects the $\rho$ width and $\rho -\omega$ mass
difference, and assumes ideal mixing and negligible flavor-breaking
in the vector meson couplings, $\rho$ and $\omega$ contributions to
the slope of $-{\frac{1}{2}}\Pi_{V\pi}+{\frac{3}{2}}\Pi_{V\eta}$
cancel exactly.  This cancellation mechanism which, owing to the near
degeneracy of the $\rho$ and $\omega$ masses, will also be present in the
vector meson contributions to the higher derivatives at $Q^2=0$, is
specific to the vector meson contributions, encoded in the NNLO and
higher LECs. There is thus no reason to expect a similar cancellation
in the corresponding NNNLO and higher loop contributions.}

We emphasize that the mass-dependent NNNLO terms considered
here further supplement the mass-independent NNNLO contribution $C^r Q^4$
added to Eq.~(\ref{Pi1}) in Refs.~\cite{GMP17,strategy}. The latter
was required to account for the deviation between the $Q^2$ dependence
of the full vacuum polarization and the NNLO ChPT expression, visible
already beyond $Q^2\approx 0.1$~GeV$^2$. Such a mass-independent
term is, of course, also present at NNNLO, but does not contribute to the
values of the slopes at $Q^2=0$ considered above.

\section{\label{discussion} Discussion}
We determined the value of the NNLO LEC $C_{93}$ from ALEPH data
for the $V$ hadronic $ud$ and $us$ spectral functions. The difference
between these two determinations gives an estimate of the systematic
uncertainty due to effects beyond NNLO in ChPT, and turns out to
dominate the total uncertainty.

One would expect that the value $C^{ud}_{93}$ is closer to the true
mass-independent result than $C^{us}_{93}$ since the pion mass is much
smaller than the kaon mass. Assuming a mass-dependent contamination
linearly dependent on the square of the meson mass, this would lead to an
extrapolated value $C_{93}^r=-0.0158$~GeV$^2$.  Such an extrapolation, however, does not take into
account the effect of the $1/N_c$-suppressed NNNLO contribution proportional
to $\d C_{93}^{(2)}$, or other higher-order effects. To be conservative
in our assessment, we therefore take as our central result for $C_{93}^r$
the value of $C^{ud}_{93}$ given in Eq.~(\ref{C93ud}) of Sec.~\ref{C93}, and assign
to this an uncertainty equal to the difference $C^{us}_{93}-C^{ud}_{93}$
(\seef\ Eq.~(\ref{DeltaC93}) in Sec.~\ref{residual}). This represents our best
estimate of the uncertainty associated with the presence of residual
higher-order mass-dependent effects. Our final result is then
\begin{equation}
\label{final}
C^r_{93}(\m=770\ \mbox{GeV})=-0.015(5)\ \mbox{GeV}^{-2}\ .
\end{equation}

It is interesting to compare the results obtained above with
estimates based on VMD. VMD leads to the expectation
$C^{ij}_{93}\sim -{\frac{f_{{\rm EM},V}^2}{4m_V^2}}$ \cite{ABT}, with
$m_V=m_\r=775$~MeV, $f_{{\rm EM},V}=f_{{\rm EM},\r}\sim 0.2$ for $ij=ud$
and $m_V=m_{K^*}=892$~MeV, $f_{{\rm EM},V}=f_{{\rm EM},K^*}
\sim f_{{\rm EM},\r}$ for $ij=us$. The resulting $ij=ud$ estimate,
$C^{ud}_{93}\sim-0.017$~GeV$^{-2}$, agrees at the $\sim 10\%$ level
with the result found in Eq.~(\ref{C93ud}). For $ij=us$, VMD correctly
predicts that $|C^{us}_{93}|<|C^{ud}_{93}|$, though the magnitude
in this case agrees with the determinations of Eqs.~(\ref{C93us1}) and
(\ref{C93us2}) only at the approximately $30\%$ level. As noted already,
the VMD estimate for $C_{93}^{{\rm eff}}$, where the existence of
strong cancellations might lead one to anticipate a much larger
fractional error, in fact, works very well. The strong cancellation
does, however, raise worries about the possible impact of
neglected NNNLO and higher loop contributions.

Finally, in Sec.~\ref{muon}, we showed that the strong cancellation
produced by NNNLO residual-mass-dependent effects in the supplemented
NNLO representation of the sum of strange connected and full disconnected
contributions calls into question the accuracy with which this
sum can be represented by a supplemented NNLO ChPT form neglecting
currently unknown NNNLO and higher-order contributions. The slope of this
sum at $Q^2=0$, in particular, could receive sizeable corrections
from such contributions, significantly impacting the accuracy with which
the associated low-$Q^2$ contributions to the muon
anomalous magnetic moment can be estimated.

\vspace{3ex}
\noindent {\bf Acknowledgments}
\vspace{3ex}

MG would like to thank the Department of Physics of the Universitat
Aut\`onoma de Barcelona, and KM and SP would like to thank the Department
of Physics and Astronomy at San Francisco State University for hospitality.
This material is based on work supported by the U.S. Department of Energy,
Office of Science, Office of High Energy Physics, under Award Number
DE-FG03-92ER40711 (MG). KM is supported by a grant from the Natural
Sciences and Engineering Research Council of Canada. SP is supported by
CICYTFEDER-FPA2014-55613-P, 2014-SGR-1450 and the CERCA Program/Generalitat
de Catalunya.

\vspace{3ex}


\end{document}